\documentclass[nopacs,nokeys]{revtex4}
\usepackage{amsfonts}
\usepackage{amsmath}
\usepackage{amssymb}
\usepackage{bbm}
\usepackage[utf8]{inputenc}
\usepackage{tensor}
\usepackage{slashed}
\usepackage[centertableaux ]{ytableau}
\usepackage{hyperref}
\usepackage{natbib}
\usepackage{enumerate}
\usepackage{braket,mleftright}
\usepackage{graphicx}
\usepackage[caption=false]{subfig}
\usepackage{subfig}
\usepackage{cleveref}

\begin{document}

\title{Kinetic mixing, custodial symmetry and a lower bound on the dark $Z^{\prime}$ mass.}
 \author{ M. Napsuciale$^{(1)}$, S. Rodr\'{\i}guez$^{(2)}$,   H. Hern\'{a}ndez-Arellano$^{(1)}$ }
\address{$^{(1)}$Departamento de F\'{i}sica, Universidad de Guanajuato, Lomas del Campestre 103, Fraccionamiento
Lomas del Campestre, Le\'on, Guanajuato, M\'exico, 37150.}
\address{$^{(2)}$Facultad de Ciencias F\'isico-Matem\'aticas,
  Universidad Aut\'onoma de Coahuila, Edificio A, Unidad
  Camporredondo, 25000, Saltillo, Coahuila, M\'exico.} 


\begin{abstract}
In this work we consider the extension of the standard model by dark fields with an Abelian $U(1)_{d}$ spontaneously broken 
gauge symmetry in a hidden dark matter scenario. Considering all the dimension four gauge invariant terms we show that the 
tree-level relation $M^{2}_{W}=M^{2}_{\tilde Z} \cos^{2} \tilde \theta_{w}$ holds and permits to 
 write the mixing angle induced by the kinetic mixing in the neutral massive gauge boson sector, $\theta_{\zeta}$, in terms of the 
 values of $M_{Z}$, the weak mixing angle and of the mass of the physical dark gauge $Z^{\prime}$ boson. At the loop level, a similar 
 relation is obtained in the $\overline{MS}$ scheme. Using the result extracted from the 
global fit to electroweak precision data for the ratio $\rho_{0}=M^{2}_{W}/\hat{c}^{2}_{Z} M^{2}_{Z}\hat{\rho}$, we obtain 
a lower bound  $M_{Z^{\prime}}> M_{Z}$ for the dark $Z^{\prime}$ mass at the $94\%$ confidence level. 
We argue that this lower bound holds in the general case of theories for physics beyond the standard model with an extra $U(1)$ 
gauge factor subgroup, whenever the extended Higgs potential respects custodial symmetry. 
\end{abstract}
\maketitle

\section{Introduction}

The possible existence of new neutral gauge bosons has been present during decades mainly due to the prediction of string 
low energy phenomenology of the existence of the product of several $U(1)$ groups at low energies, whose specific content 
depend on the chosen path for the unification of fundamental interactions (for a review and a complete list of previous references see 
 \cite{Hewett:1988xc}, \cite{Langacker:2008yv}).  One of the low energy effects of the possible existence of new neutral bosons 
is their kinetic mixing with the standard model (SM) neutral gauge bosons \cite{Holdom:1985ag},\cite{Dienes:1996zr},\cite{Babu:1997st} 
and this possibility has recently gained renewed interest because in combination with other model-dependent mechanisms allows for 
super-weak interactions of dark matter with standard model particles. Indeed, the mystery of the interactions of dark matter with SM 
particles beyond gravity is presently one of the major challenges in high energy physics and results of the intensive experimental 
search for signals of the more popular candidates for Weakly Interacting Massive Particles (WIMPs) 
yields no trace of them \cite{Arcadi:2017kky}\cite{Roszkowski:2017nbc},  which has lead to explore alternative explanations 
for the nature of dark matter. Among these ideas, the bottom-up exploration of new neutral gauge bosons, now conceived 
as part of the mediators of dark matter gauge interactions leads naturally to the existence of kinetic mixing of the SM $U(1)_{Y}$ gauge 
boson with every abelian gauge boson in the dark gauge group $G_{D}$ \cite{ArkaniHamed:2008qn},  \cite{Baumgart:2009tn}, 
\cite{Cheung:2009qd}, \cite{Ibarra:2009bm},\cite{Hook:2010tw}, \cite{Chun:2010ve}, \cite{Mambrini:2010dq}, \cite{Mambrini:2011dw}, 
\cite{Brahmachari:2014aya}, \cite{Arguelles:2016ney}, \cite{Belanger:2017vpq}, \cite{Arcadi:2018tly}, \cite{Foot:2012ai},
\cite{Kamada:2018kmi}, \cite{Rizzo:2018ntg}, \cite{Rizzo:2018joy}, \cite{Rizzo:2018vlb},  \cite{Banerjee:2019asa},  \cite{Rueter:2019wdf},\cite{Akerib:2019diq}, \cite{Lao:2020inc}, \cite{Gehrlein:2019iwl}, \cite{Kribs:2020vyk}, \cite{Binh:2020xtf},\cite{Barnes:2020vsc}.

Our interest in the subject actually comes from the results on a new proposal for dark matter based on an unconventional 
$(1,0)\oplus (0,1)$ space-time structure (tensor dark matter in a spinor-like formalism) posed in \cite{Hernandez-Arellano:2018sen} 
(details on the formal aspects of the corresponding 
fields are given in \cite{Napsuciale:2015kua}, \cite{Gomez-Avila:2013qaa} ). This is a framework where a consistent description of the 
dark matter relic density, upper bounds for dark matter annihilation into $\bar{b}b$ , $\tau^{+}\tau^{-}$, $\mu^{+}\mu^{-}$, 
$\gamma\gamma$ and results from direct detection of dark matter is obtained. The gamma ray excess from the galactic center 
around $\omega=3 ~ GeV$ can be explained in this framework only if the tensor dark matter has a mass $M\approx M_{H}/2$ 
\cite{Hernandez-Arellano:2019qgd}. The intriguing sharp prediction $M\approx M_{H}/2$ and the fact that the leading terms are 
dimension-four lead us to explore the possibility that dark matter interactions have a gauge structure, which naturally brings 
into the scenario neutral dark gauge bosons. If the dark gauge group contains a factor $U(1)_{D}$ subgroup we have to consider 
the dimension-four term for kinetic mixing with the $U(1)_{Y}$ of the SM. The main result of the present work, however, is based 
on general assumptions and goes beyond this framework. 

The kinetic mixing causes SM neutral particles to pick up a small coupling to the dark particles. 
Additional interactions are provided by the Higgs sector, since the dark Higgs fields can mix with the SM Higgs field to yield 
complementary Higgs portals to dark matter with small couplings. The specific effective interactions depend on the setup of the dark sector. 
In the simplest case of $G_{D}=U(1)_{D}$, we have dimension-four operators for kinetic mixing and Higgs mixing to be addressed below. 
 
 In preparation for a more detailed phenomenological study of specific models for dark matter with conventional or unconventional 
 space-time structures, we present here a simple and general result for the mass of the extra neutral physical boson to be denoted 
 $Z^{\prime}$ in the following : {\it whenever the complete 
 potential for the Higgs sector of the extended theory respects the custodial symmetry of the SM Higgs sector 
 \cite{Weinberg:1975gm},\cite{Susskind:1978ms}, \cite{Sikivie:1980hm}, in extensions of the standard model containing a $U(1)$ factor 
 subgroup, the mass of the extra physical gauge
 boson is bounded from below by the value of the $\rho_{0}=M^{2}_{W}/\hat{c}^{2}_{Z}M^{2}_{Z}\hat{\rho}$ parameter extracted 
 from the fit to electroweak precision data (EWPD)}. 
 This result is obtained changing the focus from the parameters of the specific model to the consequences for the masses of the physical 
 gauge bosons dictated by the symmetries of the theory in combination with results from the fit to EWPD. 
 
 Our work is organized as follows: in the next section we derive the tree-level relation connecting the mixing angle in the massive neutral 
 sector with the $Z$, $Z^{\prime}$ masses and the ratio $M^{2}_{W}/M^{2}_{Z}\cos^{2}\theta_{w}$, which in the extended theory deviates
 from the unit value. In Section III we study the modifications of this relation at the loop level and show that results from the global fit to 
 electroweak precision data yields the lower bound $M_{Z^{\prime}}> M_{Z}$ at the $94\% $ confidence level. We discuss this 
 result on the light of the custodial symmetry of the Higgs sector. Our conclusions are given in Section IV.
 
\section{ Kinetic mixing,  hidden dark matter scenarios and custodial symmetry.}
For the sake of simplicity we consider a 
spontaneously broken $U(1)_{d}$ gauge symmetry for the dark sector, but the argument works for a larger dark gauge group $G_{d}$ 
containing an abelian $U(1)_{d}$ subgroup. In this case, the Lagrangian can be written as
\begin{equation}
{\cal L}= {\cal L}_{SM} + {\cal L}_{d} + {\cal L}_{int},
\label{Lag}
\end{equation}
where ${\cal L}_{SM}$ stands for the SM Lagrangian, ${\cal L}_{d}$ denotes the $U(1)_{d}$ gauge theory Lagrangian and  
${\cal L}_{int}$ account for the dimension four (or higher if we consider it as an effective theory) terms constructed from products of 
operators on both sides, which are invariant under the whole gauge group 
$SU(3)_{c}\otimes SU(2)_{L}\otimes U(1)_{Y}\otimes U(1)_{d}$. 

In the hidden dark matter scenario, standard model fields are 
singlets of $U(1)_{d}$ and dark fields are singlets of the SM gauge group. In this case, ${\cal L}_{int}$ must be constructed with products
of singlet operators on both sides. Concerning the SM, the lowest dimension gauge singlet operators (not necessarily Lorentz invariant) 
are the $U(1)_{Y}$ stress tensor $\tilde B^{\mu\nu}$ and the Higgs operator $\tilde\phi^{\dagger}\tilde\phi$ (hereafter we use a 
tilde on the SM fields and couplings in the extended theory to distinguish them from the fields and couplings in the SM). On the 
dark side, the lowest dimension gauge singlets are the $U(1)_{d}$ stress tensor $V^{\mu\nu}$ and the dark Higgs 
operator $\Phi^{\dagger}\Phi$. The leading terms connecting the SM and dark sectors are dimension four and given by
\begin{equation}
 {\cal L}_{int} = -\frac{\sin\chi}{2} \tilde B^{\mu\nu} V_{\mu\nu} - 2 \kappa\tilde \phi^{\dagger}\tilde\phi ~ \Phi^{\dagger}\Phi.
 \label{Lint}
\end{equation}
 There could be additional terms depending on the matter content in the dark side, e.g., if there is a dark neutrino $\nu^{d}$ 
 whose right component is a singlet of $U(1)_{d}$, then $\bar{\tilde L}\tilde\phi^{c}\nu^{d}_{R}$ is also dimension four, but these 
 terms do not affect the gauge sector considered here. The Higgs sector in Eq. (\ref{Lag}) reads
\begin{equation}
{\cal L}_{Higgs}=(D^{\mu}\tilde\phi)^{\dagger}D_{\mu}\tilde\phi  + (D^{\mu}\Phi)^{*}D_{\mu}\Phi - V(\tilde\phi,\Phi)
\label{Higgslag}
\end{equation}
with the Higgs potential 
\begin{align}
V(\tilde\phi,\Phi) &=\tilde\mu^{2}\tilde\phi^{\dagger}\tilde\phi + \tilde\lambda (\tilde\phi^{\dagger}\tilde\phi)^{2} 
+\mu_{d}^{2}\Phi^{*}\Phi + \lambda_{d} (\Phi^{*}\Phi)^{2} 
+2 \kappa \Phi^{*}\Phi \tilde\phi^{\dagger}\tilde\phi .
\label{Higgspotsmu1}
\end{align}
The covariant derivative is
\begin{equation}
D^{\mu}=\partial^{\mu} + i\tilde g T^{a}\tilde W^{a\mu}  + i\tilde{g}_{Y} \frac{Y}{2} \tilde{B}^{\mu} + i g_{d}\frac{Q_{d}}{2} V^{\mu},
\end{equation}
where $Q_{d}/2$ denotes the generator of $U(1)_{d}$. The first effect of the kinetic mixing term in Eq.(\ref{Lint}) is to produce a  non-canonical 
Lagrangian in the neutral sector.  A properly normalized Lagrangian requires a redefinition of the fields by the following 
$GL(2,\mathbb{R})$ transformation \cite{ Babu:1997st}
\begin{align}
\tilde B_{\mu}&=\bar{B}_{\mu}-\tan\chi \bar{V}_{\mu},  
\qquad  V_{\mu}=\sec\chi \bar{V}_{\mu},
\end{align}
such that the kinetic sector for the gauge bosons has a canonical form 
\begin{equation}
\mathcal{L}^{K}_{gauge}=-\frac{1}{4}( \tilde W^{a\mu\nu} \tilde W^{a}_{\mu\nu}+ \bar{B}^{\mu\nu}\bar{B}_{\mu\nu}
+\bar{V}^{\mu\nu}\bar{V}_{\mu\nu}).
\label{Klagd}
\end{equation}
Writing the covariant derivative in terms of the new fields  we obtain
\begin{equation}
D^{\mu}=\partial^{\mu} + i\tilde g T^{a}\tilde W^{a\mu}  + i\tilde{g}_{Y} \frac{Y}{2} \bar{B}^{\mu} 
+ i(g_{d}\sec\chi\frac{Q_{d}}{2} -\tilde{g}_{Y} \tan\chi \frac{Y}{2})\bar{V}^{\mu}.
\end{equation}
The last term, shows that the extra gauge boson acquires a non-vanishing hypercharge due to kinetic mixing and this term, under 
spontaneous symmetry breaking (SSB), generates mixing mass terms. 
We notice that the $\bar{B}^{\mu}$ field has the 
same coupling to SM fields as the original $\tilde B^{\mu}$ field 
and in order to maintain the $U(1)_{em}$ as an unbroken symmetry with generator $Q=T_{3}+Y/2$ we need to consider 
$\bar{B}_{\mu}$ as the part of the hypercharge field $\tilde B_{\mu}$ that mixes with $\tilde W^{3}_{\mu}$. Indeed, performing 
a rotation with the weak mixing angle $\tilde{\theta}_{w}$  
\begin{equation}
\begin{pmatrix}\bar{B}\\\tilde W_{3} \end{pmatrix}
= \begin{pmatrix}\cos\tilde\theta_{w}&\ -\sin\tilde\theta_{w} \\ \sin\tilde\theta_{w} & \cos\tilde\theta_{w} \end{pmatrix} 
\begin{pmatrix}A\\ \tilde Z  \end{pmatrix}
\end{equation}
we get 
\begin{align}
\tilde g T_{3}\tilde W_{3} + \tilde{g}_{Y} \frac{Y}{2} \bar{B} 
=e Q A  + \frac{\tilde g}{\tilde{c}_{w}}  ( T_{3} - \tilde{s}^{2}_{w} Q  ) \tilde Z ,
\end{align}
where $e=\tilde{g} \tilde{s}_{w} =\tilde{g}_{Y} \tilde{c}_{w} $,  and we use the shorthand notation 
$\tilde{s}_{w} =\sin\tilde\theta_{w}$, $\tilde{c}_{w} =\cos\tilde\theta_{w}$. 

In the unitary gauge, the spontaneously broken solutions  for the Higgs fields are 
\begin{equation}
\tilde\phi=\begin{pmatrix} 0 \\ \frac{\tilde{v}+\tilde{H}}{\sqrt{2}} \end{pmatrix}, \qquad \Phi= \frac{v_{d}+\bar{S}}{\sqrt{2}},
\end{equation} 
where $\tilde{H}$ and $\bar{S}$ have vanishing vacuum expectation values. 
The minimum conditions read
\begin{equation}
\tilde\mu^{2 }+   \tilde\lambda    \tilde{v}^{2}  + \kappa v_{d}^{2}=0, \qquad \mu_{d}^{2}+  \lambda_{d}  v_{d}^{2}  +  \kappa \tilde{v}^{2}=0 ,
\end{equation}
and the following mass Lagrangian for the gauge bosons is obtained
\begin{equation}
{\cal L}^{GB}_{mass}=M^{2}_{\tilde W}\tilde W^{+\mu}\tilde W^{-}_{\mu}
+\frac{1}{2}\begin{pmatrix}\tilde Z &\bar{V} \end{pmatrix}\begin{pmatrix} M^{2}_{\tilde Z} & 
\Delta \\ \Delta & M^{2}_{\bar{V}} \end{pmatrix} \begin{pmatrix}\tilde Z\\ \bar{V}  \end{pmatrix},
\label{lagmix}
\end{equation} 
where
\begin{align}
M^{2}_{\tilde{W}}&= \frac{\tilde{g}^{2}\tilde{v}^{2}}{4}, \label{m2W}\\
M^{2}_{\tilde{Z}}&=\frac{M^{2}_{\tilde W}}{\tilde{c}^{2}_{w}}, \label{csreltree}\\
\Delta&= \frac{M^{2}_{\tilde W}}{\tilde{c}^{2}_{w}}\tilde{s}_{w} \tan\chi, \label{Delta}\\
 M^{2}_{\bar{V}}&=M^{2}_{\tilde W} \tan^{2}\tilde\theta_{w} \tan^{2}\chi + g^{2}_{d} v_{d}^{2}\sec^{2}\chi . \label{m2Vb}
\end{align} 

The neutral massive sector is diagonalized by the following rotation
\begin{equation}
\begin{pmatrix}\tilde Z\\ \bar{V} \end{pmatrix}= \begin{pmatrix}\cos\theta_{\zeta} &-\sin\theta_{\zeta}  \\ \sin\theta_{\zeta}  & \cos\theta_{\zeta}  \end{pmatrix} 
\begin{pmatrix}Z\\\ Z'  \end{pmatrix},
\label{rot}
\end{equation}
yielding the relations
\begin{align}
M^{2}_{Z} &= M^{2}_{\tilde Z}c^{2}_{\zeta} + M^{2}_{\bar{V}}s^{2}_{\zeta} + 2 \Delta s_{\zeta} c_{\zeta},  \label{mixmz} \\
M^{2}_{Z^{\prime}} &=M^{2}_{\tilde Z}s^{2}_{\zeta} + M^{2}_{\bar{V}}c^{2}_{\zeta} - 2 \Delta s_{\zeta} c_{\zeta}, \label{mixmzp} \\
 \tan 2\theta_{\zeta}  &= \frac{2\Delta}{M^{2}_{\tilde Z} -M^{2}_{\bar{V}}} . 
 \label{tan2tz}
\end{align}
Below we will find useful the converse relations 
\begin{align}
M^{2}_{\tilde Z} &= M^{2}_{Z}c^{2}_{\zeta} + M^{2}_{Z^{\prime}} s^{2}_{\zeta}, \label{mixmzbar} \\ 
M^{2}_{\bar{V}} &= M^{2}_{Z}s^{2}_{\zeta} + M^{2}_{Z^{\prime}} c^{2}_{\zeta}, \label{mixmvbar}\\ 
\Delta&=\frac{1}{2}\sin 2\theta_{\zeta} (M^{2}_{Z} -M^{2}_{Z^{\prime}}) . \label{sin2tz}
\end{align}

In the following considerations, Eq.(\ref{mixmzbar}) will be very important because it allows us to write the mixing angle 
$\theta_{\zeta}$ in terms of the physical $Z$ and $Z^{\prime}$ masses and the non-diagonal $\tilde{Z}$ mass which in turn can be written 
in terms of measurable quantities. Explicitly
\begin{equation}
s^{2}_{\zeta}=\frac{M^{2}_{\tilde{Z}}-M^{2}_{Z}}{M^{2}_{Z^{\prime}}-M^{2}_{Z}}.
\label{s2z}
\end{equation}
Considering the whole chain of transformations we have the following relations among the 
physical fields and the original gauge fields
\begin{align}
\begin{pmatrix}\tilde B \\ \tilde W_{3} \\  V\end{pmatrix}&= \begin{pmatrix} \tilde{c}_{w}, & - \tilde{s}_{w} c_{\zeta}-\tan\chi s_{\zeta}, 
& \tilde{s}_{w}s_{\zeta}-\tan\chi c_{\zeta} \\
\tilde{s}_{w} & \tilde{c}_{w}c_{\zeta}& - \tilde{c}_{w}s_{\zeta}   \\ 
0&\sec\chi s_{\zeta} & \sec\chi c_{\zeta} \end{pmatrix} \begin{pmatrix} A \\ Z\\ Z' \end{pmatrix} .
\label{mixmat}
\end{align}
The physical gauge fields couple to the SM fields through these matrix elements. Explicitly, the covariant derivative, written in 
terms of the mass eigenstates is given by
\begin{align}
D_{\mu}&=\partial_{\mu} + i \frac{\tilde g}{\sqrt{2}} (T^{+}\tilde{W}^{+}_{\mu}+T^{-}\tilde{W}^{-}_{\mu} )  + i e Q A_{\mu}  \nonumber \\
&+ i \left[\frac{\tilde g  }{ \tilde{c}_{w}} \left( c_{\zeta} ( T_{3} - \tilde{s}^{2}_{w} Q  ) - \tilde{s}_{w} s_{\zeta} \tan\chi \frac{Y}{2}\right) 
+ g_{d} s_{\zeta} \sec\chi \frac{Q_{d}}{2} \right]Z_{\mu} \nonumber \\
& - i \left[ \frac{\tilde g  }{ \tilde{c}_{w}} \left( s_{\zeta} ( T_{3} - \tilde{s}^{2}_{w} Q  ) + \tilde{s}_{w} c_{\zeta} \tan\chi \frac{Y}{2}  \right) 
- g_{d} c_{\zeta} \sec\chi \frac{Q_{d}}{2}\right] Z^{\prime}_{\mu} , 
\label{covderphys}
\end{align}
where $T^{\pm}=T^{1}\pm i T^{2}$. 

We are interested in the connection of the parameters appearing in this covariant derivative with measured physical quantities. 
In this concern, the best measured electroweak observables are the electromagnetic fine structure constant, $\alpha$, the Fermi constant, 
$G_{F}$, and the $Z$ boson mass, $M_{Z}$, and in the rest of this section we will work out the tree level relations for the connection 
of the parameters appearing in Eq. (\ref{covderphys}) with these observables. The tree level calculation of the muon lifetime yields 
the Fermi constant $G_{F}=\tilde{g}^{2}/4\sqrt{2}M^{2}_{\tilde{W}}=1/\sqrt{2}\tilde{v}^{2}$. The value of $G_{F}$ is fixed from data, 
thus $\tilde{v}$ is the conventional Higgs vacuum expectation value denoted as  $v$ in the literature, 
$\tilde{v}^{2}=v^{2}\equiv 1/\sqrt{2}G_{F}$. Next, using $\tilde{g}=e/\tilde{s}_{w}$ to write the Fermi constant in terms of 
$\alpha$ and Eq. (\ref{csreltree}) we obtain
\begin{equation}
M^{2}_{\tilde{Z}} \tilde{s}^{2}_{w} \tilde{c}^{2}_{w}=\frac{\pi\alpha}{\sqrt{2}G_{F}}\equiv{A}.
\label{Gkmt}
\end{equation}
In order to solve this equation for the weak mixing angle we need to characterize the deviation of the physical $Z$ boson mass from
its standard model value. This can be done in terms of the following tree level rho-parameter
\begin{equation}
\rho^{t}_{0}\equiv\frac{M^{2}_{\tilde{W}}}{M^{2}_{Z}\tilde{c}^{2}_{w}}=\frac{M^{2}_{\tilde{Z}}}{M^{2}_{Z}}.
\label{rho0t}
\end{equation}
In the zero mixing case ($\sin\chi=0$) we recover the standard model  tree level value $\rho^{t}_{0}=1$. Using Eq. (\ref{rho0t}) in 
Eq. (\ref{Gkmt}) we get
\begin{equation}
 \tilde{s}^{2}_{w} \tilde{c}^{2}_{w}=\frac{A}{\rho^{t}_{0}M^{2}_{Z}},
\end{equation}
which has the solutions \cite{Altarelli:1990wt}
\begin{equation}
 \tilde{s}^{2}_{w}=\frac{1}{2}-\sqrt{\frac{1}{4}-\frac{A}{\rho^{t}_{0}M^{2}_{Z}}} ,\qquad  
 \tilde{c}^{2}_{w}= \frac{1}{2}+\sqrt{\frac{1}{4}-\frac{A}{\rho^{t}_{0}M^{2}_{Z}}} .
\end{equation}
Notice that the expression of the weak mixing angle differ by the $\rho^{t}_{0}$ factors from its expression in the standard model. 
Finally, Eq. (\ref{rho0t}) yields the following expression for the $\tilde{W}^{\pm}$ mass
\begin{equation}
M^{2}_{\tilde{W}}=\rho^{t}_{0} M^{2}_{Z} \left( \frac{1}{2}+\sqrt{\frac{1}{4}-\frac{A}{\rho^{t}_{0}M^{2}_{Z}}} \right).
\end{equation} 
We remark that the $\tilde{W}^{\pm}$ is the physical propagating field but it has a mass that differs in the $\rho^{t}_{0}$ factors
from its standard model expression in terms of $\alpha$, $G_{F}$ and $M_{Z}$.

We close this section rewriting Eq. (\ref{s2z}) in terms of  $\rho^{t}_{0}$ to obtain 
\begin{equation}
s^{2}_{\zeta}=\frac{(\rho^{t}_{0}-1)M^{2}_{Z}}{(M^{2}_{Z^{\prime}}-M^{2}_{Z})}.
\label{sztree}
\end{equation}
This tree level result, relates the value of the mixing angle $\theta_{\zeta}$ to the physical values of $M_{Z}$, $M_{Z^{\prime}}$, 
and to the tree level deviation of the SM value $\rho^{t}_{0}-1$. This relation could be used to find the allowed values of 
$M_{Z^{\prime}}$ if we had the value of $\rho^{t}_{0}$ extracted from electroweak precision data. However, the current experimental 
precision demands to incorporate radiative corrections in the extraction of the value of $M_{Z}$ ( and also of $\alpha$ and $G_{F}$) from 
experimental data, thus we need to derive the relation between $s^{2}_{\zeta}$, $M^{2}_{Z}$, $M^{2}_{Z^{\prime}}$ and the analogous 
of $\rho^{t}_{0}$ at the loop level. We carry out the corresponding analysis in the next section. 

\section{Mass Lagrangian at the loop level}

Radiative corrections leave the form of the mass Lagrangian in Eq.(\ref{lagmix}) unchanged but modify the expressions for the 
masses. In order to set the notation and to make clear the observables to be calculated, we first briefly review the extraction 
from experiments of the electromagnetic fine structure constant, $\alpha$, the Fermi constant, $G_{F}$, and the $Z$ boson 
mass, $M_{Z}$. The world average for the electromagnetic fine structure at low energies, $\alpha^{-1}=137.035999084(21)$, is 
obtained from the values extracted from measurements of the electron anomalous magnetic moment and measurements of the 
Rydberg constant and atomic masses of $^{87}R_{b}$ and $^{133}Cs$. The Fermi constant 
$G_{F}=1.1663787(6)\times 10^{-5}GeV^{-2}$ is extracted from the muon lifetime. The value $M_{Z}=91.1876\pm 0.021~ GeV$ is
determined from the $Z$ lineshape at LEP (see the review {\it Electroweak Model and Constraints on
New Physics} in \cite{Zyla:2020zbs} ). Obtaining these precise values requires to incorporate radiative corrections 
and to choose a substraction scheme in their calculation in the framework of the standard model. The values above use the 
modified minimal substraction ($\overline{MS}$) scheme. In this scheme, the value of the physical Weinberg angle at the 
scale $\mu$ is given by 
\begin{equation}
\sin\hat\theta_{W}(\mu )\equiv \frac{\hat{g}^{2}_{Y}(\mu)}{\hat{g}^{2}(\mu)+\hat{g}^{2}_{Y}(\mu)},
\end{equation}
where, following the notation in \cite{Zyla:2020zbs}, a hat is used to denote the values of the SM Weinberg angle and 
electroweak couplings in the $\overline{MS}$ scheme. The gauge boson masses in this scheme are related to other observables as
\begin{equation}
M^{2}_{W}=\frac{\pi\alpha}{\sqrt{2}G_{F}\hat{s}^{2}_{Z}(1-\Delta\hat{r}_{W})}, \qquad M^{2}_{Z}=\frac{M^{2}_W}{\hat{\rho}~\hat{c}^{2}_{Z}},
\label{MWMZphys}
\end{equation}
where $\hat{s}_{Z}\equiv \sin\hat{\theta}_{W}(M_{Z})$, $\hat{c}_{Z}\equiv \cos\hat{\theta}_{W}(M_{Z})$, and the factors 
$\Delta\hat{r}_{W}$, $\hat{\rho}$ include the radiative corrections relating $\alpha, \hat{\alpha} (M_{Z}), G_{F}, M_{W}$ 
and $M_{Z}$. In the standard model, the dominant contributions to these radiative contributions in the $\overline{MS}$ are 
due to top quark loops which turn out to be quadratic in the top quark mass. Including also the $b$ quark contributions, 
it has been shown that $\hat{\rho}= 1+\hat{\rho}_{tb}$  with the loop correction  \cite{Veltman:1994wz}
\begin{equation}
\hat{\rho}_{tb}= \frac{3G_{F}}{8\sqrt{2}\pi^{2}}\left(m^{2}_{t}+m^{2}_{b}
- 2\frac{m^{2}_{t}m^{2}_{b}}{m^{2}_{t}-m^{2}_{b}} \ln \frac{m^{2}_{t}}{m^{2}_{b}} \right).
\label{rctb}
\end{equation}
In the $\overline{MS}$, these contributions are contained in $\hat{\rho}$ but do not appear in $\Delta\hat{r}_{W}$ which, to leading 
order,  gets contributions from the running of the electromagnetic coupling only. Subdominant contributions arise from Higgs boson 
loops which yields $\hat{\rho}= 1+\hat{\rho}_{tb}+\hat{\rho}_{H}$ where
\begin{equation}
\hat{\rho}_{H}= -\frac{11G_{F} M^{2}_{Z}\hat{s}^{2}_{Z}}{24\sqrt{2}\pi^{2}}  \ln \frac{M^{2}_{H}}{M^{2}_{Z}}.
\label{rcH}
\end{equation}

Including all bosonic loops, the value $\hat{\rho}=1.01019\pm0.00009$ is obtained \cite{Zyla:2020zbs}.
Notice that from Eqs. (\ref{MWMZphys}) we get the relation
\begin{equation}
M^{2}_{Z}\hat{s}^{2}_{Z}\hat{c}^{2}_{Z}=\frac{\pi\alpha}{\sqrt{2}G_{F}\hat{\rho}(1-\Delta\hat{r}_{W})}.
\label{M2zswcw}
\end{equation}

Possible effects of physics beyond the SM are explored in the global fit to the electroweak precision data (EWPD) defining the 
$\rho_{0}$ parameter which measures the deviation of the SM value (including radiative corrections of SM fields) as \cite{Zyla:2020zbs} 
\begin{equation}
\rho_{0}\equiv \frac{M^{2}_{W}}{\hat{c}^{2}_{Z}M^{2}_{Z}\hat{\rho}} .
\label{rho0}
\end{equation}
The global fit to the electroweak precision data yields  \cite{Zyla:2020zbs} 
\begin{equation}
\rho_{0}=1.00038\pm 0.00020.
\label{rho0exp}
\end{equation}

A comparison of the predictions of the extended theory with the above experimental results, requires to work out the same observables 
in this framework, including  radiative corrections in the $\overline{MS}$ scheme. 
In our formalism, the calculation of the muon lifetime 
yields a relation analogous to the first of Eqs.(\ref{MWMZphys}) given by  
\begin{equation}
M^{2}_{\tilde{W}}=\frac{\pi\alpha}{\sqrt{2}G_{F}\tilde{s}^{2}_{Z}(1-\Delta\tilde{r}_{W})},
\label{M2Wkm}
\end{equation}
where now $\tilde{s}_{Z}\equiv \tilde{s}_{w}(M_{Z})$ and $\Delta\tilde{r}_{W}$ accounts for the radiative corrections 
in the $\overline{MS}$ scheme. 

The mass Lagrangian for the neutral sector at the loop level has a similar form to Eq.(\ref{lagmix}) with the replacements 
$M^{2}_{\tilde{Z}}\to \hat{M}^{2}_{\tilde{Z}}$, $\Delta\to \hat{\Delta}$ and $M_{\bar{V}}\to \hat{M}_{\bar{V}}$, where the hat indicates 
the value in the $\overline{MS}$ at the scale $\mu=M_{Z}$. In particular, the tree level relation in Eq.(\ref{csreltree}) is modified to 
\begin{equation}
\hat{M}^{2}_{\tilde{Z}}=\frac{M^{2}_{\tilde{W}}}{\tilde{\rho}~\tilde{c}^{2}_{Z}},
\label{csrel}
\end{equation}
with $\tilde{c}_{Z}\equiv \tilde{c}_{w}(M_{Z})$ and $\tilde{\rho}$ incorporates the radiative corrections in the $\overline{MS}$ scheme. Combining this relation with Eq.(\ref{M2Wkm}) yields
\begin{equation}
\hat{M}^{2}_{\tilde{Z}}\tilde{s}^{2}_{Z}\tilde{c}^{2}_{Z}=\frac{\pi\alpha}{\sqrt{2}G_{F}\tilde{\rho}(1-\Delta\tilde{r}_{W})}.
\label{M2zswcwkm}
\end{equation}
Comparing Eqs.(\ref{M2zswcw},\ref{M2zswcwkm}) we obtain the following relation between quantities in the extended theory and 
measured quantities
\begin{equation}
\hat{M}^{2}_{\tilde Z} \tilde{c}^{2}_{Z} \tilde {s}^{2}_{Z} \tilde{\rho}(1-\Delta\tilde{r}_{W})
=M^{2}_{Z}\hat{s}^{2}_{Z}\hat{c}^{2}_{Z}\hat{\rho}(1-\Delta\hat{r}_{W}).
\label{ZZtrelfull}
\end{equation}
Now we can calculate the observable $\rho_{0}$ in Eq.(\ref{rho0}). Notice that now $M^{2}_{W}$ and $M^{2}_{Z}$ are not related 
by the second of Eqs. (\ref{MWMZphys}) and the physical $M^{2}_{\tilde{W}}$ is instead related by the custodial symmetry to the 
non-diagonal $\tilde{M}^{2}_{Z}$ through Eq. (\ref{csrel}).  With the aid of Eqs. (\ref{csrel},\ref{ZZtrelfull}) we obtain
\begin{equation}
\rho_{0}=\frac{M^{2}_{\tilde{W}}}{\hat{c}^{2}_{Z}M^{2}_{Z}\hat{\rho}} = 
\frac{\hat{M}^{2}_{\tilde{Z}} \tilde{\rho}~\tilde{c}^{2}_{Z} }{M^{2}_{Z}\hat{\rho}\hat{c}^{2}_{Z}}
=\frac{\hat{s}^{2}_{Z}(1-\Delta\hat{r}_{W})}{\tilde{s}^{2}_{Z}(1-\Delta\tilde{r}_{W})}.
\label{rho0km0}
\end{equation}
Next we split the contributions to radiative corrections into SM contributions and contributions from new particles.  
For the SM extension by a $U(1)_{d}$ dark matter gauge group in a hidden scenario, the couplings of the dark matter 
particles (the dark $Z^{\prime}$ and dark matter fields) generated by the kinetic mixing or the mixing in the Higgs sector in 
Eq. (\ref{Lint}) are small, thus radiative corrections due to these particles can be neglected in a first approximation and we can safely 
keep only radiative corrections due to standard model particles. 
The dominant terms in the SM radiative corrections are due to the 
running of $\alpha $ and to quadratic top mass corrections. Subdominant corrections due to the Higgs mass are 
logarithmic and involves the $Z$ mass thus they will introduce small differences because the kinetic mixing shifts the value of 
$M_{Z}$. These are however small corrections in a subdominant term and it is safe to neglect them. Within these approximations 
we can take $\Delta\tilde{r}_{W}=\Delta\hat{r}_{W}$
and Eq.(\ref{rho0km0}) reduces to
\begin{equation}
\rho_{0}= \frac{\hat{s}^{2}_{Z}}{\tilde{s}^{2}_{Z}}.
\label{rho0km}
\end{equation}

The diagonalization process for the loop-level mass Lagrangian follows along the lines of the tree-level one discussed above. 
The explicit form of the loop level mass terms, $\hat\Delta$ and $\hat{M}^{2}_{\bar{V}}$, are not relevant 
for the derivation of the main result in this paper. The diagonalization yields a relation similar to Eq. (\ref{s2z}), namely
\begin{equation}
\hat{s}^{2}_{\zeta}=\frac{\hat{M}^{2}_{\tilde{Z}}-M^{2}_{Z}}{M^{2}_{Z^{\prime}}-M^{2}_{Z}},
\label{mainp}
\end{equation} 
where $\hat{s}_{\zeta}\equiv \sin\theta_{\zeta}(M_{Z})$ and $\hat{M}^{2}_{\tilde{Z}}$ is given in Eq.(\ref{csrel}) in the 
$\overline{MS}$ scheme at the scale $\mu=M_{Z}$. Using Eq. (\ref{rho0km}) it is posible to show that
\begin{equation}
\hat{M}^{2}_{\tilde{Z}}=\frac{\rho_{0}^{2}\hat{c}^{2}_{Z}}{\rho_{0}-\hat{s}^{2}_{Z}}M^{2}_{Z}.
\end{equation}
Using now this relation in Eq.(\ref{mainp}) we get
\begin{equation}
\hat{s}^{2}_{\zeta}=\frac{(\rho_{0}-1)(\rho_{0}\hat{c}^{2}_{Z} - \hat{s}^{2}_{Z})M^{2}_{Z}}{(\rho_{0} 
-\hat{s}^{2}_{Z})(M^{2}_{Z^{\prime}}-M^{2}_{Z})}.
\label{main}
\end{equation} 

Equation (\ref{main}) is the main result of this paper. It relates the physical mixing angle $\theta_{\zeta}$ to the values of 
the Weinberg angle, the $Z$ mass and $\rho_{0}-1$ at the scale $\mu=M_{Z}$ in the $\overline{MS}$ scheme. Notice that
this relation is similar to the tree-level one in Eq.(\ref{sztree}), but now all the involved quantities are the ones extracted from data, 
except for the unknown physical dark $Z^{\prime}$ mass. Also, $\rho_{0}$ includes already the radiative corrections due to 
standard model particles, thus the deviation from the SM value $\rho_{0}=1$ is due to new physics only. In our case it contains 
the leading effects of the kinetic mixing given by the tree level contributions of the Lagrangian in Eq.(\ref{Lint}).

Since the value of $\hat{s}^{2}_{\zeta}$ is restricted to $0\leq \hat{s}^{2}_{\zeta}\leq 1$, 
the result of the fit to electroweak precision data in Eq. (\ref{rho0exp}) and Eq.(\ref{main}) yields the following constraint for 
the dark $Z^{\prime}$ mass
\begin{equation}
 M_{Z^{\prime}} \geq M_{Z} \sqrt{1+ \frac{(\rho_{0}-1)(\rho_{0}\hat{c}^{2}_{Z} - \hat{s}^{2}_{Z})}{(\rho_{0} -\hat{s}^{2}_{Z})}},
\label{lbd}
\end{equation}
a result valid at the $1.9~\sigma$ level ($94\%$ confidence level) for the dark $Z^{\prime}$ mass. The second term in the square root
of this equation is small and positive thus we can simply state the lower bound as $M_{Z^\prime}>M_{Z}$.

The lower bound in Eq. (\ref{lbd}) has been derived for a $U(1)_{d}$ extension of the SM in a hidden scenario. 
However, it is well known that the tree-level relation  $M^{2}_{\tilde Z}=M^{2}_{\tilde W}/\tilde c^{2}_{w}$ has its root in the 
$SO(4)\sim SU(2)_{L}\otimes SU(2)_{R}$ global symmetry of the SM Higgs sector which under SSB is broken down to 
its $SU(2)_{V}$ subgroup, dubbed custodial symmetry in \cite{Sikivie:1980hm}.  
The $SU(2)_L$ gauge bosons transforms in the triplet representation of this remnant symmetry which requires the 
triplet to have a common mass, thus the mass term for the $\tilde W^{\mu}_{3}$ is given as 
$M^{2}_{\tilde W_{3}}=M^{2}_{\tilde W_{\pm}}$. The non-vanishing vacuum expectation value of the Higgs field generates a 
non-diagonal mass term given by $g g_{Y}\langle \phi_{0}|T_{3}Y|\phi_{0}\rangle \tilde{W}^{\mu}_{3}\bar{B}_{\mu}$, which 
introduces the weak mixing angle to yield the tree level relation 
$M^{2}_{\tilde{W}_3}=M^{2}_{\tilde Z}\tilde c^{2}_{w}=M^{2}_{\tilde W^{\pm}}$. 
Custodial symmetry protects this relation from large radiative corrections. At tree level, non-diagonal matrix elements vanish 
in the $g_{Y}\to 0$ limit.  Also bosonic loop corrections vanish in the $g_{Y}\to 0$ limit, whilst fermionic loop corrections vanish 
in the limit when up-type and down-type quarks have the same mass as it is obvious from Eq. (\ref{rctb}). In general, custodial symmetry is 
broken by the hypercharge and Yukawa couplings but modifications of the tree level relation in Eq.(\ref{csreltree}) 
by radiative corrections are small  and yields  Eq.(\ref{csrel}) with $\tilde{\rho}\approx \hat{\rho}$ at the electroweak scale.
If we go beyond the hidden scenario and provide the field $\tilde\phi$ with a dark charge and the dark Higgs $\Phi$ with a 
non-vanishing hypercharge, new mass mixing terms and additional contributions to the $\bar{V}$ mass terms will be generated
but, whenever the dark field has vanishing electric charge and weak isospin, the custodial symmetry protected relation
$M^{2}_{\tilde Z}\tilde c^{2}_{w}=M^{2}_{\tilde{W}}$ still holds and the lower bound in Eq. (\ref{lbd}) still applies.

\section{ Conclusions}

In this work we focus on an extension of the standard model by spontaneously broken Abelian $U(1)_{d}$ dark matter gauge theory 
in a hidden dark matter scenario. In this case, the tree level relation relation $M^{2}_{\tilde{W}}=M^{2}_{\tilde Z} \cos^{2}\tilde\theta_{w}$ 
makes possible to rewrite the mixing angle induced by the kinetic mixing in the neutral massive gauge sector, $\theta_{\zeta}$, in 
terms of the tree-level  
values of $M_{Z}$, $\theta_{w}$ and the mass of the physical new dark gauge boson $M_{Z{^\prime}}$. Then we go through the 
loop level analysis since the present-day precision in the extraction of the values of  $M_{Z}$ and $\theta_{w}$ requires to incorporate
radiative corrections. We argue that for the considered extension of the standard model, radiative corrections are dominated by 
standard model particles and obtain the corresponding expression for the mixing angle in terms of the values of 
$M_{Z}$, $\theta_{w}(M_{Z})$ and $M_{Z^{\prime}}$ in the $\overline{MS}$ at the scale $\mu=M_{Z}$.
We use this relation and the value of $\rho^{0}\equiv M^{2}_{W}/\hat{c}^{2}_{Z} M^{2}_{Z}\hat\rho$  extracted from the global 
fit to electroweak precision data, where $\hat{c}_{Z}=cos\theta_{w}(M_{Z})$ and $\hat{\rho}$ denote the radiative corrections 
due to standard model particles, to obtain the lower bound $M_{Z^{\prime}}> M_{Z}$. We recall that the tree-level relation 
$M^{2}_{\tilde{W}}=M^{2}_{\tilde Z} \cos^{2}\tilde\theta_{w}$ is due to the custodial 
symmetry of the standard model Higgs potential and show that in gauge theories for physics beyond the standard model with gauge groups 
containing a new $U(1)$ factor group, whenever this symmetry is respected, the outlined procedure can be carried out and the mass of the physical neutral extra gauge boson still satisfies the lower bound $M_{Z^{\prime}}> M_{Z}$. On the light of our calculations, the possibility 
of a light dark $Z^{\prime}$ requires to consider extensions of the standard model with non-Abelian dark gauge symmetry groups 
containing no $U(1)$ factor groups.

\section{Acknowledgments}
One of us (H.H.A.) acknowledges CONACyT-M\'{e}xico for a scholarship to pursue her Ph. D. . 

\bibliographystyle{JHEP}
\bibliography{kineticmixing}
\end{document}